
%
%
\tolerance 1000
\magnification 1200
\baselineskip=18pt plus 1pt minus 1pt
\def\vs{{\bf S}}
\newcount\eqnumber
\eqnumber=0
\def\en{\global\advance\eqnumber by 1
          \eqno(\the\eqnumber)}
\def\ket#1{\vert#1\rangle}

\hsize=6.25 truein
\vsize=8.5 truein
\pageno=1

\vglue 0.5 truein
\centerline{\bf Exact Results of Dimerization Order Parameter in}
\centerline{\bf $SU(n)$ Antiferromagnetic Chains}
\vskip .4 truein \centerline{Y. Xian}
\vskip .4 truein \centerline{\sl Department of Mathematics}
\centerline{\sl University of Manchester Institute of Science and
                Technology (UMIST)}
\centerline{\sl P.O. Box 88, Manchester M60 1QD, England}
\vskip.4truein

\noindent{\bf Abstract}: After a proper definition of the dimerization
order parameter for a spin-$S$ system, I show that this order
parameter in the $SU(n)$ ($n=2S+1$) antiferromagnetic chains (or
equivalently the $SU(2)$ spin-$S$ chains with Hamiltonians which
project out singlet states) is, in the thermodynamic limit, directly
proportional to the staggered-magnetization in the corresponding
spin-$1\over2$ $XXZ$ chains which had already been mapped onto the
$SU(n)$ chains.

\vskip 1 truein
\noindent PACS numbers: 75.10.Jm, 75.40.Cx

\vfill
\eject

The spin-1 antiferromagnetic chain with the pure biquadratic exchange
has the Hamiltonian
$$ H = -\sum_i (\vs_i\cdot\vs_{i+1})^2, \en$$
where the summation over $i$ runs over all spins with either free
ends or the usual periodic boundary condition. Parkinson [1] first
discussed the possibility of a mapping of Eq.~(1) onto the
spin-$1\over2$ $XXZ$ chain with the anisotropy $\Delta = {3\over2}$, which
is in general, apart from a constant, described by the Hamiltonian
$$ H = -\sum_i H_{i\,i+1}^{xxz}, \ \ \ \ H^{xxz}_{i\,i+1} \equiv
   {1\over2}(\sigma^x_i\sigma^x_{i+1} + \sigma^y_i\sigma^y_{i+1}) +
   {1\over2}\Delta (1- \sigma^z_i\sigma^z_{i+1}),\en$$
where $\sigma^\alpha\ (\alpha =x,y,z)$ are Pauli matrices and $\Delta$
is the anisotropy parameter. Barber and Batchelor [2] later have shown
that the Hamiltonian of Eq.~(1) with free ends is indeed exactly
equivalent to the 9-state quantum Potts chain. They then obtained the
ground-state energy and the excitation gap by the mapping of the Potts
chain onto the spin-$1\over2$ chains of Eq.~(2) with $\Delta
={3\over2}$, and with fields $\pm{1\over4}\sqrt5$ applied to the two
ends respectively. Kl\"umper [3] obtained independently these exact results,
and he also presented results for the correlation length.

Based on these exact results, Kl\"umper [3] and Affleck [4] showed
that the ground state and low-lying excited states of a series of
$SU(n)$ (with $n = 2S+1$) antiferromagnetic chains can all be
similarly obtained. In particular, Affleck [4] showed that the $SU(n)$
chains with free ends can be mapped, in a similar fashion, onto the
corresponding spin-$1\over2$ $XXZ$ chains of Eq.~(2). The generic
Hamiltonian of these $SU(n)$ chains with free ends are given by
$$ H = -\sum_{i=1}^{N-1} P^0_{i\,i+1}(S), \en $$
where $N$ is the number of spins in the chain, and $P^J_{i\,j}(S)$ is
the projection operator which projects out the state with total spin
$J$ of the pair $\vs_i$ and $\vs_j$ with $(\vs_i+\vs_j)^2 = J(J+1)$.
For $S={1\over2}$, Eq.~(3) reduces to the usual Heisenberg model plus
a constant, whereas for $S=1$, Eq.~(3) is equivalent to the pure
biquadratic chain given by Eq.~(1) with free ends because
$P^0_{i\,i+1} = [(\vs_i\cdot\vs_{i+1})^2 -1]/3$. For $S={3\over2}$,
Eq.~(3) becomes,
$$ H = {1\over1152}\sum_{i=1}^{N-1} [60(\vs_i\cdot\vs_{i+1})^3
       +80(\vs_i\cdot\vs_{i+1})^2-372\vs_i\cdot\vs_{i+1}-297].\en$$
We note that this form of the Hamiltonian is quite similar to that of
the spin-$3\over2$ chain proposed by Babujian [5], which is fully
integrable by Bethe's {\it ansatz}. The mapping of Eq.~(3) for a
general $n =2S+1$ onto the spin-$1\over2$ chains of Eq.~(2) is given
by the relation, $ n =2 \Delta$. In this paper, I shall use the
free-end boundary condition as in Ref. [2] and [4], and with even
total number of spins $N$ in the $N \rightarrow \infty$ limit. The
final results are independent of the boundary condition in the
thermodynamic limit.

Although the exact values of the ground-state energy and excitation
gap have been obtained for the Hamiltonians of Eqs.~(1) and (3) for a
general $n$ by the mapping onto the spin-$1\over2$ $XXZ$ chains of
Eq.~(2) which was exactly solved by Bethe's {\it ansatz}, and although
people are convinced [2,4] that the systems described by Eq.~(3) are
dimerized for any $n > 2$, it does not seem possible to calculate
directly the dimerization order parameter [6] which is usually
defined in the thermodynamic limit, by
$$ D \equiv \langle(\vs_{i-1}\cdot\vs_i - \vs_i\cdot\vs_{i+1})\rangle,
             \en$$
where the angular brackets denote a ground-state expectation. We
note that the absolute value of $D$ is independent of $i$ in the
thermodynamic limit (but $i$ should be far away from boundaries since the
free-end boundary condition is used here).

In the course of studying spin-lattice dimerization and trimerization
problems [7], I have come to realize that there is another effective,
perhaps more proper, definition of the dimerization order parameter
for a general spin-$S$ system. This new order parameter, which was
given by the ground-state expectation of a square matrix with
dimension given by the number of states for a two-atom spin-$S$ system
[7], can in fact be equivalently expressed as the ground-state
expectation of the projection operator $P^0_{i\,j}$ of Eq.~(3) as,
$$ D(n) \equiv \langle(P^0_{i-1\,i}(S)-P^0_{i\,i+1}(S))\rangle, \ \ \ \
        n=2S+1. \en$$
For $S={1\over2}$, this definition of order parameter is
identical to the usual definition of Eq.~(5), but it is certainly
different from Eq.~(5) for any $S > {1\over2}$, though it is obvious
that both definitions can effectively measure the order of
dimerization. It is clear that the key difference between the
definitions of Eqs.~(5) and (6) lies in the fact that operator
$\vs_i\cdot\vs_{i+1}$ of Eq.~(5) has in principle a projection of all
states with a $J$ value of the paired spins with
$(\vs_i+\vs_{i+1})^2=J(J+1)$, unlike $P^0_{i\,i+1}$ of Eq.~(6), which
projects out only the singlet state of the pair.

Both Eqs.~(5) and (6) for the definition of dimerization order
parameter are still meaningful for systems with the periodic boundary
condition. However, some care should be taken since a dimerized system
has two degenerate ground states and the expectation with respect to
an equal admixture of them will yield zero result in Eqs.~(5) and (6).
With the free-end boundary condition, one has the advantage of a
non-degenerate ground state.

To see how one comes to the definition of Eq.~(6), we consider the
case of perfect dimerization for a spin-$S$ chain.  It is convenient
to discuss dimerized states in the valence-bond basis. Spin operators
can be usefully written in terms of two pairs of Schwinger bosons as,
$$ S^+ = a^+b,\ \ \ \ S^-=ab^+,\ \ \ \ S^z={1\over2}(a^+a-b^+b), \en$$
where $a,a^+$ and $b,b^+$ obey the usual boson commutation relations.
In this representation, a spin-$S$ state with $S^z=m\ (-S\le m \le S)$
is written as
$$\ket m = {(a^+)^{S+m}\over\sqrt{(S+m)!}}
{(b^+)^{S-m}\over\sqrt{(S-m)!}} \ket 0,\en $$
where $\ket 0$ is the vacuum state of the bosons.  A valence bond is
simply a spin-singlet configuration, which can be written by the so
called valence bond operator $C_{i\,j}^+$, defined by
$$ C_{i\,j}^+ \equiv a_i^+b_j^+ - a_j^+b_i^+.\en$$
For example, the singlet state of two-atom
spin-$1\over2$ system is given by a single valence-bond configuration,
$$ \ket{\Phi_1} = C_{i\,j}^+\ket 0 = \ket{{1\over2},-{1\over2}} -
\ket{-{1\over2},{1\over2}}; \en$$
for $S=1$, this singlet state is
given by a two-bond configuration,
$$ \ket{\Phi_2}= (C_{i\,j}^+)^2\ket
0 = 2(\ket{1,-1} + \ket{-1,1} - \ket{0,0}).  \en $$
Generally, the singlet ground state of a two-atom system, each with
spin $S$, is given by a $2S$-bond configuration as
$$
\ket{\Phi_{2S}}=(C^+_{i\,j})^{2S}\ket 0.\en$$ The perfect dimerization
state, $\ket{D}$, of the spin-$S$ chain can then be written as $$
\ket{D} = \left({1\over n[(n-1)!]^2}\right)^{N\over4}
\prod_{i=1}^{N/2} (C^+_{2i-1\,2i})^{n-1} \ket 0, \ \ \ \ n=2S+1,\en$$
where I have included the normalization factor. It is clear that if
the periodic boundary condition is used, the dimerization state of
Eq.~(13) will be doubly degenerate, as mentioned earlier. But as we
are using the free-end boundary condition here, $\ket D$ of Eq.~(13)
is the only choice for the perfect dimer state. The basic algebras in
the $SU(n)$ chain of Eq.~(3) are given by the following two
operations,
$$ P^0_{i\,j}(S)\, (C_{i\,j}^+)^{2S}\ket 0 = (C_{i\,j}^+)^{2S}\ket
       0; \en$$
and
$$ P^0_{i\,j}(S)\, (C_{k\,i}^+)^{2S}(C_{j\,l}^+)^{2S}\ket 0 =
   {1\over n}\, (C_{i\,j}^+)^{2S}(C_{l\,k}^+)^{2S}\ket 0, \en$$
where the four indices $k,i,j,l$ are all different from one another
and, as before, $n=2S+1$. From these two equations, one can in fact
prove that the operator, $n*P^0_{i\,i+1}(S)$, obeys the Temperley-Lieb
algebra [8], which is the key to the mapping of Eq.~(3) onto the
spin-$1\over2$ $XXZ$ chain of Eq.~(2) with the free-end boundary
condition [2,4], by considering all possible valence-bond
configurations of the type of Eq.~(12) involving four consecutive
atoms. Affleck [4] has provided another more elegant proof by using
the fermion representation.

One can also define a normalized version of Eq.~(6) as
$$ D'(n) \equiv {1\over D_0}\langle(P^0_{i-1\,i}(S)
   - P^0_{i\,i+1}(S))\rangle, \ \ \ \ n=2S+1,\en$$
where $D_0$ is the expectation value of $(P^0_{i-1\,i}(S) -
P^0_{i\,i+1}(S))$ with respect to the perfect dimerized state $\ket
{D}$ of Eq.~(13).  Using Eqs.~(13)-(15), it is a straightforward
calculation to obtain this expectation value as $D_0 = 1 - 1/n^2$.

Since operator $n*P^0_{i\,i+1}(S)$ obeys the Temperley-Lieb algebra [2,4],
one can write
$$   P^0_{i\,i+1} \rightarrow {1\over n} \bigl[H^{xxz}_{i\,i+1}
    +{1\over2}\sinh\theta\,(\sigma^z_{i+1}-\sigma^z_i)\big], \en$$
where $\sigma^\alpha\ (\alpha = x,y,z)$ are Pauli matrices and
$H^{xxz}_{i\,i+1}$ is given by Eq.~(2) with $\Delta = \cosh\theta =
{n \over 2}$.  Using this transformation,
one can straightforwardly calculate the order parameter $D'(n)$ of
Eq.~(16) by using the mappings of the exact ground-state of Eq.~(3)
onto that of the corresponding spin-$1\over2$ chains of Eq.~(2).
One thus obtains
$$ D'(n) = {n\over n^2-1}\,\sqrt{n^2-4}\,\langle\sigma_i^z\rangle,\en$$
where the expectation is with respect to the ground-state of the
spin-$1\over2$ $XXZ$ chain of Eq.~(2) with the anisotropy $\Delta = n/2$, and
where I have used the fact that, {\it in the thermodynamic limit}, one
has
$$ \langle H^{xxz}_{i-1\,i}\rangle = \langle H^{xxz}_{i\,i+1}\rangle, \ \ \ \
   {\rm and}\ \ \langle\sigma_{i-1}^z\rangle = \langle\sigma_{i+1}^z\rangle
    =-\langle\sigma_i^z\rangle, \en$$
in the same expectation. Eq.~(19) simply reflects the well-known fact
that the infinite spin-$1\over2$ $XXZ$ chain has no dimerization
long-range order but a staggered magnetization order. Fortunately,
this staggered magnetization had already been exactly calculated by
Baxter [9] twenty years ago as a function of the anisotropy,
$$ \eqalign{
   \sqrt{\langle\sigma_i^z\rangle} &= 1+2\sum_{n=1}^\infty(-1)^n
       \exp(-2n^2\theta)\cr
        &=\sqrt{2\pi\over\theta}\sum_{n=1}^\infty
      \exp[-(n-1/2)^2\pi^2/2\theta],\cr}\en$$
where two expressions are equivalent, the first being rapidly
convergent at large $\theta$ while the second at small $\theta$. For
$S=1$ and $3\over2$, $\langle\sigma_i^z\rangle\approx 0.5028$ and $0.7335$
respectively to the accuracy of four significant figures. Therefore,
one has $D'(3) \approx 0.4216$ and $D'(4) \approx 0.6776$ to the same
accuracy of four significant figures for the corresponding $SU(3)$ and
$SU(4)$ chains of Eq.~(3) respectively. One sees also that in the
limit of $S \rightarrow \infty$, $\langle\sigma_i^z\rangle = 1$ and
hence $D'(\infty)=1$ as expected.

Based on the definition of the order parameter in Eq.~(6), one can
define a corresponding `four-spin' correlation function as $\langle
(P^0_{i\,i+1} P^0_{j\,j+1}) \rangle$, in similar fashion to the usual
definition of the four-spin correlation function, $\langle
(\vs_i\cdot\vs_{i+1}) (\vs_j\cdot\vs_{j+1}) \rangle$, for the order
parameter of Eq.~(5). Likewise, by taking the long-range limit (i.e.,
$\vert j-i \vert \rightarrow \infty$) in $\langle (P^0_{i\,i+1}
P^0_{j\,j+1}) \rangle$, one should be able to obtain the value of
order parameter $D(n)$ (or $D'(n)$). This is useful if one is to carry
out finite-size calculations with periodic boundary conditions.

\vskip 1.5 truein

\centerline{\bf Acknowledgements}
\vskip .3 truein
I thank R.F. Bishop, C.E. Campbell, and J.B. Parkinson for many useful
discussions.

\vfill
\eject

\centerline{\bf References}
\vskip .3 truein

\item{[1]} J.B. Parkinson, J. Phys. C: Solid State Phys. {\bf
20} (1987) L1029; J. Phys. C: Solid State Phys. {\bf 21} (1988) 3793.

\item{[2]} M.N. Barber and M.T. Batchelor, Phys. Rev. B {\bf
40} (1989) 4621.

\item{[3]} A. Kl\"umper, Europhys. Lett. {\bf 9} (1989) 815.

\item{[4]} I. Affleck, J. Phys.: Condens. Matter {\bf 2} (1990) 405.

\item{[5]} H.M. Babujian, Phys. Lett. A {\bf 90} (1982) 479.

\item{[6]} F.D.M. Haldane, Phys. Rev. B {\bf 25} (1982) 4925; A.V.
Chubukov, Phys. Rev. B {\bf 43} (1991) 3337.

\item{[7]} Y. Xian, J. Phys.: Condens. Matter {\bf 5} (1993) 7489.

\item{[8]} H.N.V. Temperley and E.H. Lieb, Proc. R. Soc.
A {\bf 322} (1971) 251; R.J. Baxter, {\it Exactly Solved Models in
Statistical Mechanics} (New York: Academic, 1982).

\item{[9]} R.J. Baxter, J. Stat. Phys. {\bf 9} (1973) 145.

\vfill \bye